# Anisotropic charged dark energy star

Kanika Das[1] and Nawsad Ali[2]

*Department of Mathematics,*

*Gauhati University, Guwahati- 781014, Assam, India*

*Email: [1]daskanika2@gmail.com; [2] alinawsad@rocketmail.com*

**Abstract:** As the stars carry electrical charges, we present in this paper a model for charged dark energy star which is singularity free. We take Krori-Barua space time. We assume that the radial pressure exerted on the system due to the presence of dark energy is proportional to the isotropic perfect fluid matter density and the difference between tangential and radial pressure is proportional to the square of the electric field intensity. The solution satisfies the physical conditions inside the star.

**Keywords:** Dark energy, Singularity-free, Einstein-Maxwell equations, anisotropic stresses.

## 1. Introduction

Recent cosmological observations point towards an accelerated expansion of the universe at the present epoch. The astronomical observations of the present universe provide evidence for the existence of a kind of energy called dark energy which governs the expansion of the universe [1-3]. The most recent WMAP observations indicate that dark energy accounts for 72% of total mass energy of the universe [4]. However, the nature of dark energy is still a mystery. Many cosmologists believe that the best candidate for the dark energy is the cosmological constant $\Lambda$, which is usually interpreted physically as a vacuum energy with energy density $\rho_\Lambda$ and pressure $p_\Lambda$ satisfying the usual equation of state (EOS) $\rho_\Lambda = -p_\Lambda$. However, one has the reason to dislike the cosmological constant since it always suffers the theoretical problems such as the "fine-tuning" and "cosmic coincidence" puzzles [5]. That is why, the different forms of dynamically changing dark energy with an effective equation of state $\omega = \frac{p}{\rho} < -\frac{1}{3}$, have been proposed in literature.

A large amount of work on charged fluid spheres is available. With regard to singularity we would like to mention that Efinger [6], Kyle and Martin [7] and Wilson [8] have found relativistic internal solutions for static charged spheres, but none of these solutions is absolutely free from singularities. It is observed that in Efinger's solution the metric has a singularity at the region $(r = 0)$ whereas the solutions to Kyle and Martin and Wilson do not have singularities at $r = 0$. But in both the cases the metrices may have singularities at points other than the origin so that restrictions have to be imposed on the sphere to avoid them. The fluid sphere solution of Kyle and Martin, Wilson, Kramer and Neugebauer [9] and Krori and Barua [10] are of special interest since with the imposition of suitable condition, they are completely free of metric singularities and satisfy physical conditions. Krori and Barua [10] constructed static spherically symmetric solutions of the Einstein-Maxwell equations based on particular choice of the metric components $g_{00}$ and $g_{11}$ in curvature coordinates. A thorough analysis of this singularity-free (KB) [10] solution has been done by Junevicus [11]. According to Ivanov [12], the presence of charge function serves as a safety valve, which absorbs much of the fine-tuning necessary in the uncharged case. However, in connection to this we would like to mention a special kind of mechanism to avert singularity as used by Trautman [13] under Einstein-Cartan theory. In recent times cosmological models based on dark energy have attained significant attention. Rahaman, et al. [14] have found relativistic internal solutions for dark energy star. Varela et al. [15], have dealt with self-gravitating, charged, anisotropic fluids and got more flexibility in solving Einstein-Maxwell equations. In order to discuss analytical solutions they extend Krori-Barua's method to include pressure anisotropy and linear and non-linear equations of state. Also Rahaman et al. [16] have extended Krori-Barua analysis of the static, spherically symmetric, Einstein-Maxwell field equations and have considered charged fluid sources with anisotropic stresses. They have included a new variable (tangential pressure) which allows the use of a nonlinear, Chaplygin-type equation of state with coefficients fixed by the matching conditions at the boundary of the source. Moreover, Rahaman et al. [17] have investigated the ultra compact objects like strange stars in Krori-Barua space time.

Rahaman et al. [15] have obtained a model for singularity-free dark energy star. As stars carry electrical charges we propose, in this paper, a model for an anisotropic charged dark energy star. Here we assume that the radial pressure exerted on the system due to the presence of dark energy is proportional to the isotropic perfect fluid matter density and the difference between tangential and radial pressures is proportional to the square of electric field intensity. The stellar configuration comprises charge, an ordinary baryonic perfect fluid together with an unknown form of matter (dark energy) which is repulsive in nature. We also assume that the two fluids are non-interacting amongst themselves with the radial and transverse directional anisotropic property such that $p_r \neq p_t$ [18]. Recent observations on highly compact astrophysical objects like X ray pulsar Her X-1, X ray buster 4U 1820-30, millisecond pulsar SAX J 1808, 4-3658, X-34, etc., also strongly favor an anisotropic matter distribution since the density inside such an ultra compact object is expected to be beyond nuclear matter density.

The scheme of the study is as follows: After introduction in section 1, we present the field equations and their solutions in connection with the proposed model for charged dark energy star in section 2. In section 3-7 are dealt, respectively, with the boundary conditions, TOV equation, energy conditions, stability, and mass radius relation. Finally the paper ends with conclusion in section 8.

## 2. The field equations and their solutions

We take the Krori-Barua [10] metric given by

$$ds^2 = -e^{\nu(r)}dt^2 + e^{\lambda(r)}dr^2 + r^2 d\theta^2 - r^2 \sin^2\theta\, d\phi^2 , \qquad (1)$$

with $\lambda = Ar^2$ and $\nu = Br^2 + C$, where A, B and C are arbitrary constants which will be fixed on the ground of various physical requirements.

We take the energy momentum tensor for anisotropic charged with two fluids as

$$T_0^0 = (\rho)_{eff} = \rho + \rho_{de} + E^2, \tag{2}$$

$$T_1^1 = -(p_r)_{eff} = -(p + p_{der}) - E^2, \tag{3}$$

$$T_2^2 = T_3^3 = -(p_t)_{eff} = -(p + p_{\mathrm{det}}) + E^2, \tag{4}$$

where $\rho$, $p$ and $E$ correspond to the energy density, pressure of the baryonic matter and electric field intensity, respectively, and $\rho_{de}$, $p_{der}$ and $p_{\mathrm{det}}$ are the 'dark' energy density, radial pressure and tangential pressure, respectively.

The Einstein-Maxwell field equations are

$$8\pi(\rho + \rho_{de}) + E^2 = e^{-\lambda}\left(\frac{\lambda'}{r} - \frac{1}{r^2}\right) + \frac{1}{r^2}, \tag{5}$$

$$8\pi(p + p_{der}) - E^2 = e^{-\lambda}\left(\frac{\nu'}{r} + \frac{1}{r^2}\right) - \frac{1}{r^2}, \tag{6}$$

$$8\pi(p + p_{\mathrm{det}}) + E^2 = e^{-\lambda}\left(\frac{\nu''}{2} + \frac{\nu'^2}{4} - \frac{\lambda'\nu'}{4} + \frac{\nu' - \lambda'}{2r}\right), \tag{7}$$

and

$$\left(r^2 E\right)' = 4\pi r^2 \sigma e^{\lambda/2}. \tag{8}$$

Equation (8) can equivalently be expressed in the form

$$E(r) = \frac{1}{r^2}\int_0^r 4\pi r^2 \sigma e^{\lambda/2} dr = \frac{q(r)}{r^2}, \tag{9}$$

where $\sigma = \sigma(r)$ is the charge density and $q(r)$ is the total charge of the sphere under consideration.

Now we consider the KB ansatz:

$$\lambda = Ar^2\,,\ \nu = Br^2 + C\,, \tag{10}$$

where, as mentioned earlier, $A$, $B$, and $C$ are some arbitrary constants to be determined on physical grounds. Using (10) in equations (5)-(7) we obtain

$$8\pi(\rho + \rho_{de}) + E^2 = e^{-Ar^2}\left(2A - \frac{1}{r^2}\right) + \frac{1}{r^2}\,, \tag{11}$$

$$8\pi(p + p_{der}) - E^2 = e^{-Ar^2}\left(2B + \frac{1}{r^2}\right) - \frac{1}{r^2}\,, \tag{12}$$

$$8\pi(p + p_{\det}) + E^2 = e^{-Ar^2}\left[(2B + A) + B(B - A)r^2\right]. \tag{13}$$

One now has at hand three equations, namely, the Einstein-Maxwell equations (11)-(13) with six unknown functions of $r$ i.e., $\rho$, $\rho_{de}$, $p$, $p_{der}$, $p_{\det}$ and $E^2$. Obtaining explicit solutions to the above field equations is extremely difficult to the nonlinearity of the equations, although the problem is mathematically well defined. However, in the spirit of Ref. [18] and [19], three assumptions are adopted. Firstly, we assume that the dark energy radial pressure is proportional to the dark energy density, i.e.

$$p_{der} = -\rho_{de}\,, \tag{14}$$

secondly, the variable dark energy density is proportional to the mass density, i.e.,

$$\rho_{de} = \alpha\rho\,, \tag{15}$$

where $\alpha > 0$ is a proportionality constant, and thirdly, the difference of tangential and radial pressure is proportional to the square of the electric field intensity i.e.,

$$p_{\det} - p_{der} = \frac{E^2}{4\pi}\,. \tag{16}$$

Using (14)-(16) in equations (11)-(13) we get the expressions for $\rho$, $p$ and $E^2$ as

$$8\pi\rho = \frac{e^{-Ar^2}}{4(1+\alpha)}\left[9A - B(B-A)r^2 - \frac{3}{r^2}\right] + \frac{3}{4(1+\alpha)r^2}\,, \tag{17}$$

$$8\pi p = \frac{e^{-Ar^3}}{4(1+\alpha)}\left[8B(1+\alpha) + B(B-A)r^2 + A(8\alpha-1) + \frac{3}{r^2}\right] - \frac{3}{4(1+\alpha)r^2}\,, \tag{18}$$

$$E^2 = \frac{e^{-Ar^2}}{4}\left[B(B-A)r^2 - A - \frac{1}{r^2}\right] + \frac{1}{4r^2}\,, \tag{19}$$

The charge density is obtained as

$$\sigma = \frac{e^{-\frac{Ar^2}{2}}}{4\pi r}\psi + \frac{e^{-\frac{3Ar^2}{2}}}{8\pi\psi r}\left[AB(A-B)r^4 + \left\{A^2 + B(B-A)\right\}r^2 + A\right] + \frac{e^{-\frac{Ar^2}{2}}}{8\pi\psi r^3}\left\{e^{-Ar^2} - 1\right\}, \tag{20}$$

where

$$\psi = \sqrt{e^{-Ar^2}\left[B(B-A)r^2 - A - \frac{1}{r^2}\right] + \frac{1}{r^2}}\,, \tag{21}$$

and the charge within a sphere of radius $r$ turns out to be

$$q = \frac{r^2}{2}\sqrt{e^{-Ar^2}\left[B(B-A)r^2 - A - \frac{1}{r^2}\right] + \frac{1}{r^2}}\,. \tag{22}$$

Thus the effective energy density $(\rho)_{eff}$, effective radial pressure $(p_r)_{eff}$ and the effective tangential pressure $(p_t)_{eff}$ are obtained as

$$(\rho)_{eff} = \frac{e^{-Ar^2}}{32\pi}\left[9A - (B-A)r^2 - \frac{3}{r^2}\right] + \frac{3}{32\pi r^2}\,, \tag{23}$$

$$(p_r)_{eff} = \frac{e^{-Ar^2}}{32\pi}\left[8B + B(B-A)r^2 - A + \frac{3}{r^2}\right] - \frac{3}{32\pi r^2}\,, \tag{24}$$

$$(p_t)_{eff} = \frac{e^{-Ar^2}}{32\pi}\left[8B + 3B(B-A)r^2 - 3A + \frac{1}{r^2}\right] - \frac{1}{32\pi r^2}\ . \quad (25)$$

Using equations (23)-(25), the equation of state (EOS) corresponding to radial and transverse directions may be written as

$$\omega_r(r) = \frac{e^{-Ar^2}\left[8B + B(B-A)r^2 - A + \frac{3}{r^2}\right] - \frac{3}{r^2}}{e^{-Ar^2}\left[9A - B(B-A)r^2 - \frac{3}{r^2}\right] + \frac{3}{r^2}}\ . \quad (26)$$

$$\omega_t(r) = \frac{e^{-Ar^2}\left[8B + 3B(B-A)r^2 - 3A + \frac{1}{r^2}\right] - \frac{1}{r^2}}{e^{-Ar^2}\left[9A - B(B-A)r^2 - \frac{3}{r^2}\right] + \frac{3}{r^2}}\ . \quad (27)$$

It is interesting to note that the electric field intensity $(E)$, proper charge density $(\sigma)$, total charge density $(q)$, effective energy density $(\rho_{eff})$, effective radial pressure $(p_{r\,eff})$ and effective tangential pressure $(p_{t\,eff})$ are independent of $\alpha$.

We also have

$$\frac{d\rho_{eff}}{dr} = -\frac{1}{16\pi}\left[e^{-Ar^2}\left\{\left(9A^2 + B(B-A)\right)r - AB(B-A)r^3 - \frac{3A}{r} - \frac{3}{r^3}\right\} + \frac{3}{r^2}\right] < 0\,, \quad (28)$$

and

$$\frac{dp_r}{dr} = -\frac{1}{16\pi}\left[e^{-Ar^2}\left\{\left(8AB - A^2 - B(B-A)\right)r + AB(B-A)r^3 + \frac{3A}{r} + \frac{3}{r^3}\right\} - \frac{3}{r^2}\right] < 0\,. \quad (29)$$

Note that, at $r = 0$, our model provides

$$\frac{d\rho_{eff}}{dr} = 0\,,\ \frac{dp_{r\,eff}}{dr} = 0\,, \quad (30)$$

$$\frac{d^2\rho_{eff}}{dr^2} = -\frac{1}{16\pi}\left(15A^2 + B(B-A)\right) < 0\,, \quad (31)$$

and

$$\frac{d^2 p_{r\,eff}}{dr^2} = -\frac{1}{16\pi}\left(8AB - 7A^2 + B(A-B)\right) < 0\,, \tag{32}$$

Therefore, at the centre, we also see that the effective radial pressure is maximum and it decreases from the centre towards the boundary. Thus, the effective energy density and effective radial pressure are well behaved in the interior of the stellar structure. Variation of energy density and two pressures have been shown in Fig. 1 and Fig. 2, respectively.

It is interesting to note that, the bound on the effective EOS in this construction is given by $0 < \omega_i(r) < 1$ (see Fig. 3) despite of the fact that star is constituted by the combination of ordinary matter, dark energy and effect of charge.

The measure of anisotropy, $\Delta = \frac{2}{r}(p_{t\,eff} - p_{r\,eff})$ in this model is obtained as

$$\Delta = \frac{1}{8\pi r}\left[e^{-Ar^2}\left\{B(B-A)r^2 - A - \frac{1}{r^2}\right\} + \frac{1}{r^2}\right]. \tag{33}$$

This anisotropy will be directed outward when $p_t > p_r$ i.e. $\Delta > 0$, and inward if $p_t < p_r$ i.e. $\Delta < 0$. It is apparent from the Fig. 4 of our model with a repulsive 'anisotropic' force ($\Delta > 0$) allows the construction of more massive distributions.

**3. Boundary Conditions**

The exterior space-time of the star is described by the Reissner-Nordstrom metric [20, 21]

$$ds^2 = -\left(1 - \frac{2M}{r} + \frac{Q^2}{r^2}\right)dt^2 + \left(1 - \frac{2M}{r} + \frac{Q^2}{r^2}\right)^{-1} dr^2 + r^2 d\theta^2 + r^2 \sin^2\theta\, d\phi^2\,, \tag{34}$$

where $Q$ is the total charge up to the boundary surface $r = r_1$. We assume the continuity of the metric functions $g_{tt}$, $g_{rr}$ and $\frac{\partial g_{tt}}{\partial r}$ across the boundary surface S at $r = r_1$. This gives us the following equations

$$1-\frac{2M}{r_1}+\frac{Q^2}{r_1{}^2}=e^{Br_1{}^2+C}\,, \tag{35}$$

$$1-\frac{2M}{r_1}+\frac{Q^2}{r_1{}^2}=e^{-Ar_1{}^2}\,, \tag{36}$$

$$\frac{M}{r_1{}^2}-\frac{Q^2}{r_1{}^3}=Br_1e^{Br_1{}^2+C}\,. \tag{37}$$

From equations (35)-(37) one can easily get

$$A=-\frac{1}{r_1{}^2}\log\left(1-\frac{2M}{r_1}+\frac{Q^2}{r_1{}^2}\right). \tag{38}$$

$$B=\frac{1}{r_1{}^2}\left(\frac{M}{r_1}-\frac{Q^2}{r_1{}^2}\right)\left(1-\frac{2M}{r_1}+\frac{Q^2}{r_1{}^2}\right)^{-1}. \tag{39}$$

$$C=\log\left(1-\frac{2M}{r_1}+\frac{Q^2}{r_1{}^2}\right)-\left(\frac{M}{r_1}-\frac{Q^2}{r_1{}^2}\right)\left(1-\frac{2M}{r_1}+\frac{Q^2}{r_1{}^2}\right)^{-1}. \tag{40}$$

We also impose the boundary conditions that at the boundary $(p_r)_{eff}(r=r_1)=0$ and $\rho_{eff}(r=0)=b$ ($=a$ constant), where $b$ is the central density. Thus,

$$A=\frac{8\pi b}{3}\,, \tag{41}$$

$$B=\frac{\frac{8\pi b}{3}r_1{}^2-8\pm\sqrt{\left(8-\frac{8\pi b}{3}r_1{}^2\right)^2+4r_1{}^2\left\{\frac{8\pi b}{3}+\frac{3}{r_1{}^2}\left(e^{\frac{8\pi b}{3}r_1{}^2}-1\right)\right\}}}{2r_1{}^2}\,. \tag{42}$$

Combining (38) and (41), we get

$$A=\frac{8\pi b}{3}=-\frac{1}{r_1{}^2}\log\left(1-\frac{2M}{r_1}+\frac{Q^2}{r_1{}^2}\right). \tag{43}$$

At this juncture, to get an insight of our model, we find the numerical values of the parameters $A$, $B$ and $b$ for the charged dark energy star-4U 1820-30, Her X-1 and SAX J 1808.4-3658 (see Table1).

## 4. TOV Equation

For an anisotropic fluid distribution, the generalized TOV equation has the form

$$\frac{d(p_{r\,eff})}{dr}+\nu'(\rho_{eff}+p_{r\,eff})+\frac{2}{r}(p_{t\,eff}-p_{r\,eff})=0\,. \tag{44}$$

Following Ponce de Leon [23], we write the above TOV equation as

$$-\frac{M_G(\rho_{eff}+p_{r\,eff})}{r^2}e^{\frac{\lambda-\nu}{2}}-\frac{dp_{r\,eff}}{dr}+\sigma\frac{q}{r^2}e^{\frac{\lambda}{2}}+\frac{2}{r}(p_{t\,eff}-p_{r\,eff})=0\,, \tag{45}$$

where $M_G=M_G(r)$ is the effective gravitational mass inside a sphere of radius $r$ and is given by

$$M_G(r)=\frac{1}{2}r^2e^{\frac{\nu-\lambda}{2}}\nu'=Br^3e^{\frac{1}{2}[(B-A)r^2-C]}\,. \tag{46}$$

This can be derived from the Tolman-Whittaker formula and the Einstein-Maxwell's field equations. The modified TOV equation describes the equilibrium condition for a charged dark energy star subject to gravitational ($F_g$), hydrostatic ($F_h$), electric ($F_e$) and another force term arising out of anisotropy $(F_a)$ so that

$$F_a+F_h+F_e+F_a=0\,, \tag{47}$$

where

$$F_g\equiv -Br(\rho_{eff}+p_{r\,eff})=-\frac{Br}{4\pi}(A+B)e^{-Ar^2}\,, \tag{48}$$

$$F_h\equiv-\frac{dp_{r\,eff}}{dr}=\frac{e^{-Ar^2}}{16\pi}\left[\left\{8AB-A^2-B(B-A)\right\}r+AB(B-A)r^3+\frac{3A}{r}+\frac{3}{r^3}\right]-\frac{3}{16\pi r^3}\,, \tag{49}$$

$$F_e \equiv \sigma E e^{\frac{Ar^2}{2}} = \frac{1}{8\pi r}\left[e^{-Ar^2}\left\{B(B-A)r^2 - A - \frac{1}{r^2}\right\} + \frac{1}{r^2}\right] + \frac{(e^{-Ar^2}-1)}{16\pi r^3}$$

$$+\frac{e^{-Ar^2}}{16\pi r}\left[-AB(B-A)r^4 + \left\{A^2 + B(B-A)\right\}r^2 + A\right] \tag{50}$$

$$F_a \equiv \frac{2}{r}(p_{t\,eff} - p_{r\,eff}) = \frac{e^{-Ar^2}}{8\pi}\left[B(B-A)r - \frac{A}{r} - \frac{1}{r^3}\right] + \frac{1}{8\pi r^3}. \tag{51}$$

The profile of $F_g$, $F_h$, $F_e$ and $F_a$ for our chosen source are showing in Fig. 6. This figure indicates that the static equilibrium can be attained due to the combined effect of pressure anisotropy, gravitational, electric and hydrostatic forces.

## 5. Energy conditions

It is well known for the charged fluid that the null energy condition (NEC), the weak energy condition (WEC), the strong energy condition (SEC) and the dominant energy condition (DEC) will be satisfied if and only if the following inequalities hold simultaneously at every point within the source:

$$\rho_{eff} + \frac{E^2}{8\pi} \geq 0, \tag{52}$$

$$\rho_{eff} + p_{r\,eff} \geq 0, \tag{53}$$

$$\rho_{eff} + p_{t\,eff} + \frac{E^2}{4\pi} \geq 0, \tag{54}$$

$$\rho_{eff} + p_{r\,eff} + 2p_{t\,eff} + \frac{E^2}{4\pi} \geq 0, \tag{55}$$

Direct plotting of the left sides of (52)-(55) shows that these inequalities are satisfied as well as at every $r$ (see Fig. 7).

The anisotropy, as expected, vanishes at the centre i.e., $p_{t\,eff} = p_{r\,eff} = p_{0\,eff} = \frac{2B-A}{8\pi}$ at $r=0$.

Also, at $r=0$, $E=0$ and for $r=8km$ and $\frac{M}{R}=0.5$, we find from equation (22) as $q^2 = 0.198145$. Thus our model satisfies Ponce de Leon's condition [22]. The effective energy density and the two pressures are also well behaved in the interior of the stellar configuration.

Now, the energy conditions [23] at the centre can be written as follows:

(i) NEC: $p_{0\,eff} + \rho_{0\,eff} \geq 0 \Rightarrow A+B \geq 0$.

(ii) WEC: $p_{0eff} + \rho_{0eff} \geq 0 \Rightarrow A+B \geq 0$ and $\rho_{0eff} \geq 0 \Rightarrow A \geq 0$.

(iii) .SEC: $p_{0eff} + \rho_{0eff} \geq 0 \Rightarrow A+B \geq 0$ and $3p_{0eff} + \rho_{0eff} \geq 0 \Rightarrow B \geq 0$.

(iv) DEC: $\rho_{0eff} > \left| p_{0eff} \right| \Rightarrow 2A \geq B \geq 0$.

**6. Stability**

The velocity of sound $v_s^2 = \left( \frac{dp}{d\rho} \right)$ should be less than one for a realistic model [24, 25]. Now, we calculate the radial and transverse speed for our anisotropic model,

$$v_{sr}^2 = \frac{dp_{r\,eff}}{d\rho_{eff}} = -1 + \frac{8A(A+B)re^{-Ar^2}}{\left\{ 9A^2 + B(B-A) - AB(B-A)r^2 - \frac{3A}{r^2} \right\} re^{-Ar^2} + \frac{3}{r^3}\left(1-e^{-Ar^2}\right)}, \tag{56}$$

$$v_{st}^2 = \frac{dp_{t\,eff}}{d\rho_{eff}} = \frac{e^{-Ar^2}\left[ (11AB - 3A^2 - 3B^2)r + 3AB(B-A)r^3 + \frac{A}{r} + \frac{1}{r^3} \right] - \frac{1}{r^3}}{\left\{ 9A^2 + B(B-A) - AB(B-A)r^2 - \frac{3A}{r^2} \right\} re^{-Ar^2} + \frac{3}{r^3}\left(1-e^{-Ar^2}\right)}, \tag{57}$$

To check whether the sound speeds lie between $0$ and $1$, we plot the radial and transverse sound speeds in Fig. 8 and observe that these parameters satisfy the inequalities $0 \leq v_{sr}^2 \leq 1$ and $0 \leq v_{st}^2 \leq 1$ everywhere within the stellar object.

Equations (56) and (57) lead to

$$v_{st}^2 - v_{sr}^2 = 1 - \frac{e^{-Ar^2}\left[(3AB - 11A^2 - 3B^2)r + 3AB(B-A)r^3 + \frac{A}{r} + \frac{1}{r^3}\right] - \frac{1}{r^3}}{\left\{9A^2 + B(B-A) - AB(B-A)r^2 - \frac{3A}{r^2}\right\} re^{-Ar^2} + \frac{3}{r^3}\left(1 - e^{-Ar^2}\right)}. \qquad (58)$$

From equation (58), we note that $v_{st}^2 - v_{sr}^2 \leq 1$. Since, $0 \leq v_{sr}^2 \leq 1$ and $0 \leq v_{st}^2 \leq 1$, therefore, $\left|v_{st}^2 - v_{sr}^2\right| \leq 1$. In Fig. 9, we have plotted $\left|v_{st}^2 - v_{sr}^2\right|$.

A few years back, Herrera [24] proposed a technique for stability check of local anisotropy matter distribution. This technique is known as the cracking (or overturning) concept which states that the region for which radial speed of sound is greater than the transverse speed of sound is a potentially stable region. In our case, Fig. 10 indicates that there is no change of sign for the term $v_{st}^2 - v_{sr}^2$ within the specific configuration. Also, the plot for $v_{st}^2 - v_{sr}^2$ (Fig. 10) shows negative in its nature. Therefore, we conclude that charged dark energy star model is stable.

## 7. Mass-radius relation

In this section, we study the maximum allowable mass-radius ratio in our model. For a static spherically symmetric perfect fluid star, Buchdahl [26] showed that the maximally allowable mass-radius ratio as $\frac{2M}{R} < \frac{8}{9}$ which has later been generalized by Mak et al. [27] for a charged sphere. In our model, the effective generalized mass is obtained as

$$M_{eff} = 4\pi \int_0^R \left(\rho + \rho_{de} + \frac{E^2}{8\pi}\right) r^2 dr. \qquad (59)$$

In Fig. 11, we plot this mass-radius relation. We have also plotted $\frac{M_{eff}}{r_1}$ against $r_1$ (see Fig. 12) which shows that the ratio $\frac{M_{eff}}{r_1}$ is an increasing function of the radial parameter. We note that a constraint on the maximum allowable mass-radius ratio in our case is similar to that of an isotropic fluid sphere, i.e. $\frac{mass}{radius} < \frac{4}{9}$, found by Buchdahl [26]. If we define the compactification factor as

$$u = \frac{M_{eff}(r_1)}{r_1} = \frac{1}{2}\left(1 - e^{-Ar_1^2}\right), \tag{60}$$

the surface redshift ($Z_s$) corresponding to the above compactness ($u$) is obtained as

$$Z_s = (1 - 2u)^{-\frac{1}{2}} - 1, \tag{61}$$

where

$$Z_s = e^{\frac{A}{2}r_1^2} - 1. \tag{62}$$

Thus, the maximum surface redshift for an anisotropic star of radius 8 km turns out to be $Z_s = 0.07723$ (see Fig. 13).

Similarly, a lower bound on the mass to radius ratio for a charged sphere has been reported recently by Andreasson [28] which has the form

$$\sqrt{M} < \frac{\sqrt{R}}{3} + \sqrt{\frac{R}{9} + \frac{Q^2}{3R}}. \tag{63}$$

This inequality is applicable to any model satisfying the inequality $p_{r\,eff} + 2p_{t\,eff} \le \rho_{eff}$. In Fig. 14, we have plotted $p_{r\,eff} + 2p_{t\,eff} - \rho_{eff}$ against $r$ which shows that in the region $1.5 < r \le 8$, $p_{r\,eff} + 2p_{t\,eff} - \rho_{eff}$ is negative. Since our model is stable within $1.5 < r \le 8$, Andreasson's relation also holds true in our model.

## 8. Discussions

In this paper, we have found the astrophysical relevance of a charged dark energy stellar model. We have taken KB model [10] in the modeling of charged *dark* energy star. Dark energy stellar models have found astrophysical relevance for various reasons, one particular reason being its importance as an alternative candidate to a black hole. As the stars carry electrical charges, we present in this paper a model for charged dark energy star which is singularity free. We have extended KB approach assuming singularity-free form of the metric ansatz to charged anisotropic source. From equations (2) and (3) we have seen, the bound on the effective EOS in this construction is given by $0 < \omega_i(r) < 1$, (see Fig. 3) though the star is constituted by the combination of charge, ordinary matter and dark energy.

The matching of interior metric with exterior (Reissner-Nordstrom) metric, firstly, assumed continuity of the metric functions $g_{tt}$, $g_{rr}$ and $\frac{\partial g_{tt}}{\partial r}$ at the boundary surface $S$, secondly imposed the boundary conditions that at the boundary $\left(p_r\right)_{eff}(r = r_1) = 0$ and $\rho_{eff}(r = 0) = b$ ($=$a constant), where $b$ is the central density. Based on the analytic model developed so far, to get an estimate of the range of various physical parameters, let us now consider some charge star candidates like neutron stars, white dwarf stars, 4U 1820-30, Her X-1, SAX J1808.4-3658. In spirit of Ref. [10, 14, 15, 16, 17] both $A$ and $B$ are positive. From equations (42) and (43) assuming the estimated mass and radius of these stars, we have calculated the values of the relevant physical parameters as compiled in Table 1 which confirmed validity of the present model. Thus our approach leads to a better analytical description of the charged dark energy star.

Table 1 Values of parameters for different charge dark energy stars

| **Charged dark energy star** | $M(M_{\otimes})$ | $R$ **(km)** | $\frac{M}{R}$ | $A$ **(km$^{-2}$)** | $B$ **(km$^{-2}$)** | $b$ **(km$^{-2}$)** |
|---|---|---|---|---|---|---|
| **Her X-I** | **0.88** | **7.7** | **0.168** | **0.00690381** | **0.004196608** | **0.0008245** |
| **SAX J 1808.4-3658(SS1)** | **1.435** | **7.07** | **0.299** | **0.0183209** | **0.013916673** | **0.0021880** |
| **4U 1820-30** | **2.25** | **10.0** | **0.332** | **0.011026** | **0.00913881** | **0.0013168** |

## Acknowledgements

The authors express their profound gratitude to Prof.K.D.Krori for his encouragement and help and the authors acknowledge the financial support of UGC, New Delhi and the Department of Mathematics, Gauhati University for providing all facilities for doing this work.

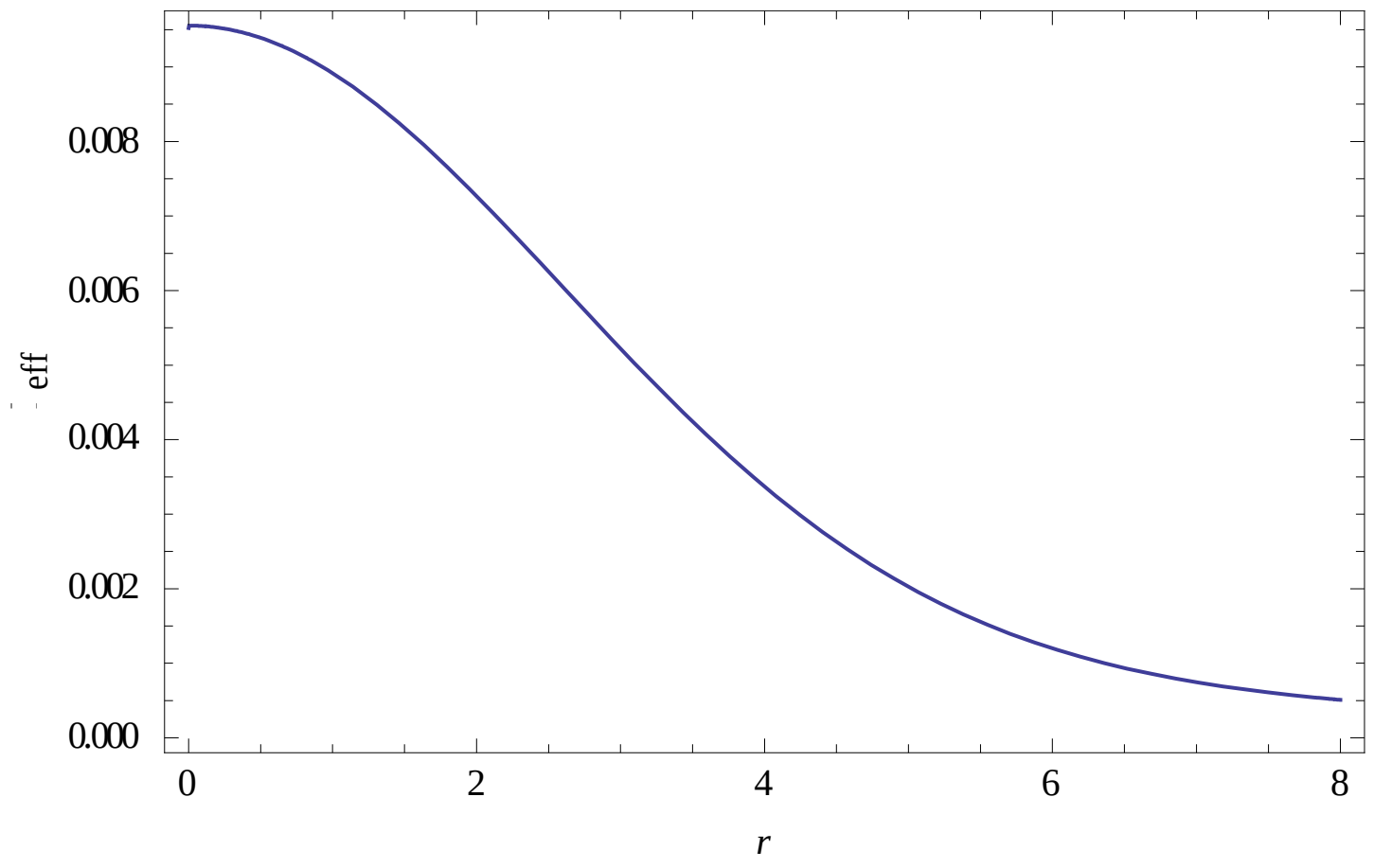

Fig. 1: The effective density parameter $\rho_{eff}$ is shown against $r$ .by taking $A = 0.0183209$ , $B = 0.01391667\ 3$ and $\frac{M}{R} = 0.299$ .

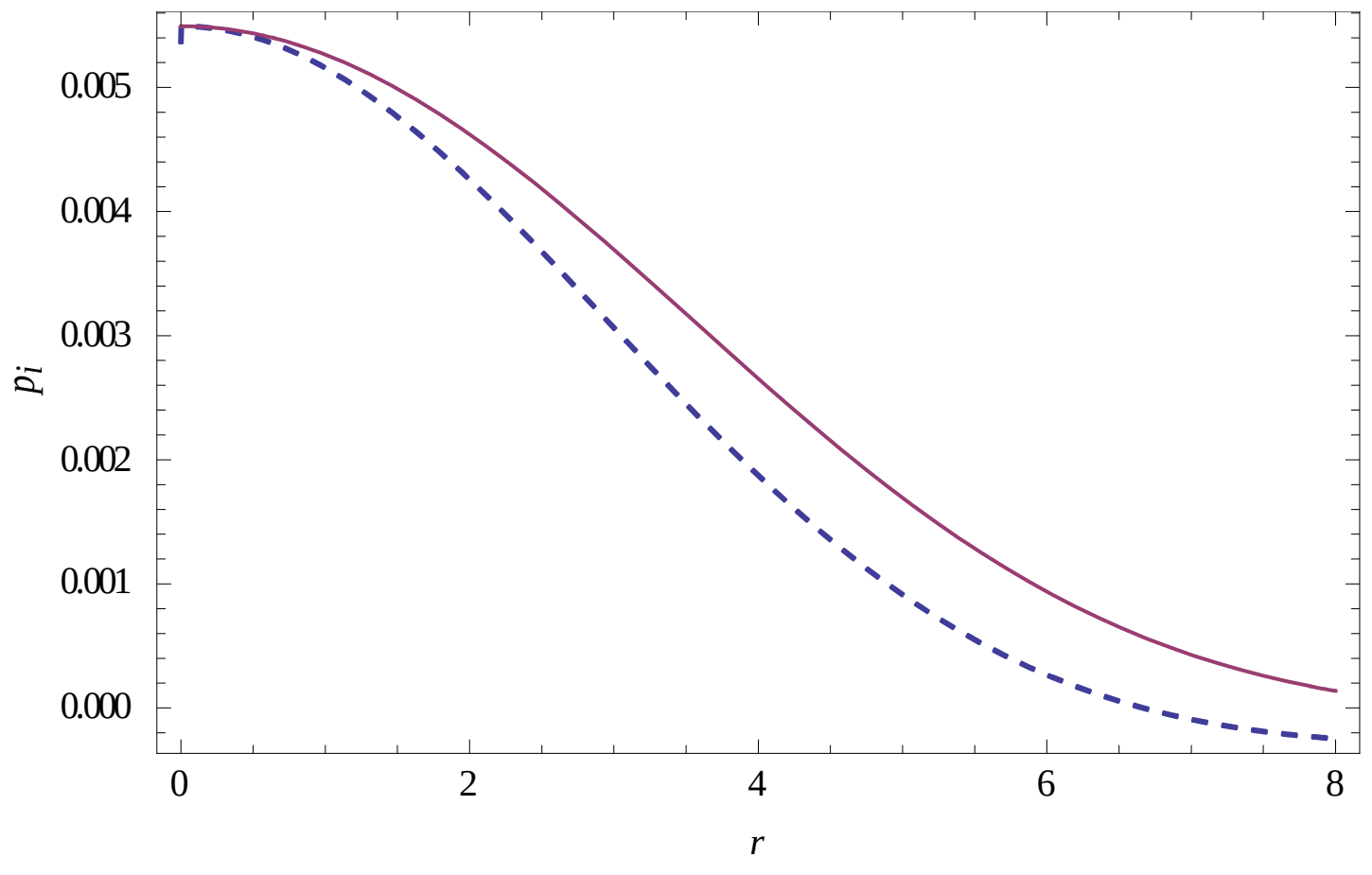

Fig.2: The dotted and solid line represents effective radial and transverse pressures respectively against $r$ by taking $A = 0.0183209$ , $B = 0.013916673$ and $\frac{M}{R} = 0.299$ .

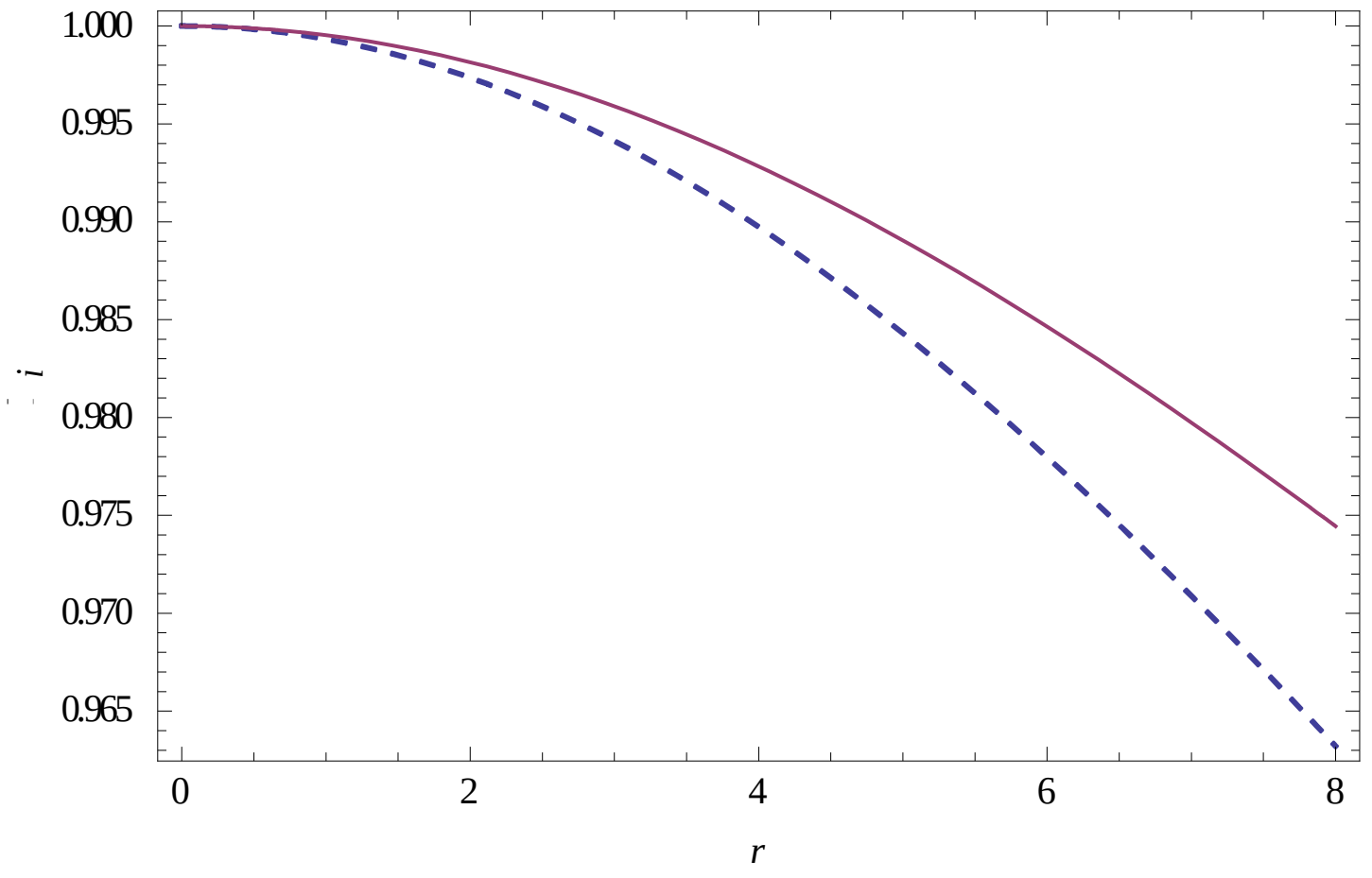

Fig.3: The dotted and solid line represents effective radial and transverse equation of state parameters respectively against $r$ by taking $A = 0.0183209$ , $B = 0.013916673$ and $\frac{M}{R} = 0.299$ .

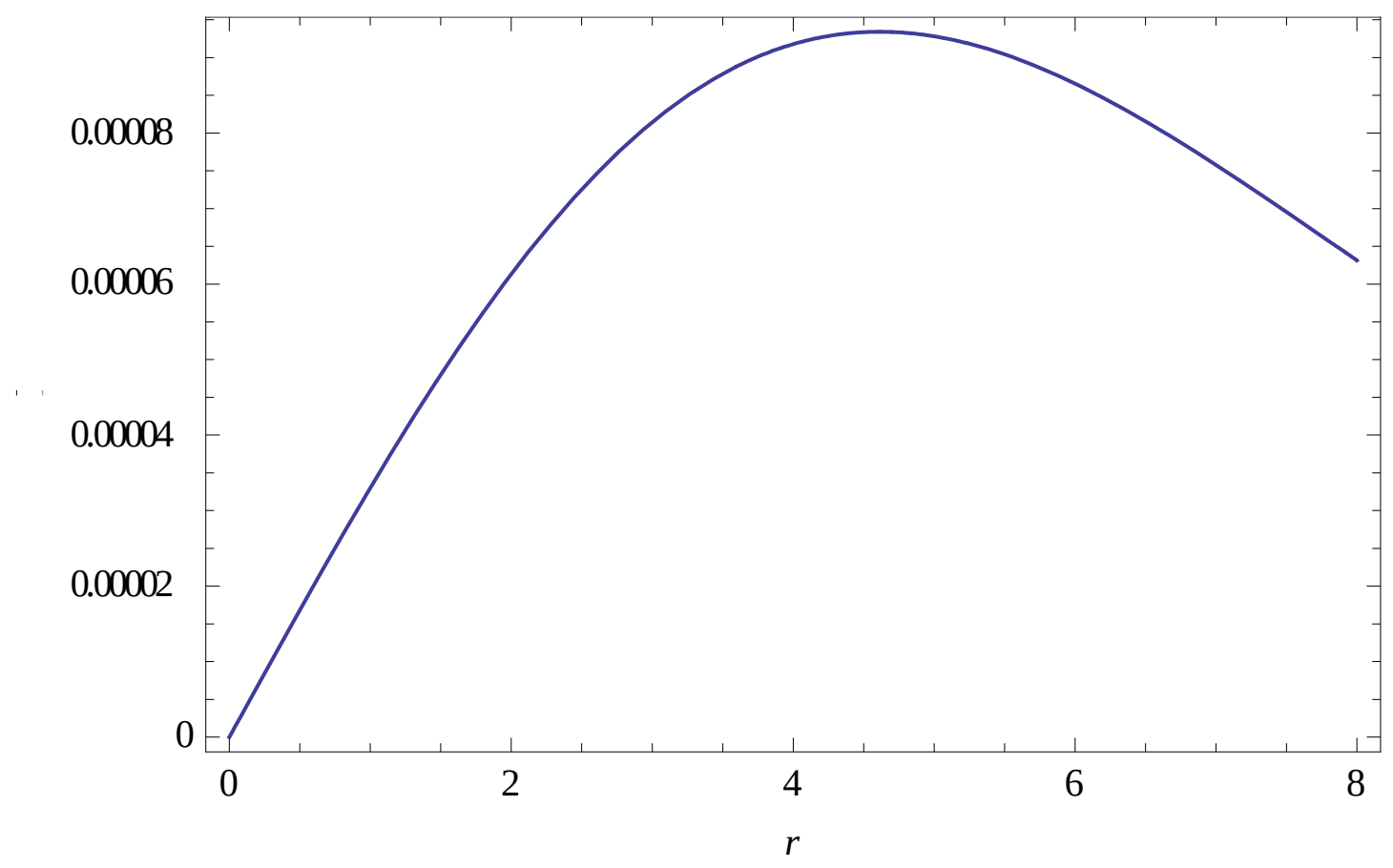

Fig. 4: The variation of the force, $\Delta = \frac{2}{r}(p_{t\,eff} - p_{r\,eff})$ due to the local anisotropy with respect to $r$ by taking $A = 0.0183209$ , $B = 0.0139166673$ and $\frac{M}{R} = 0.299$ .

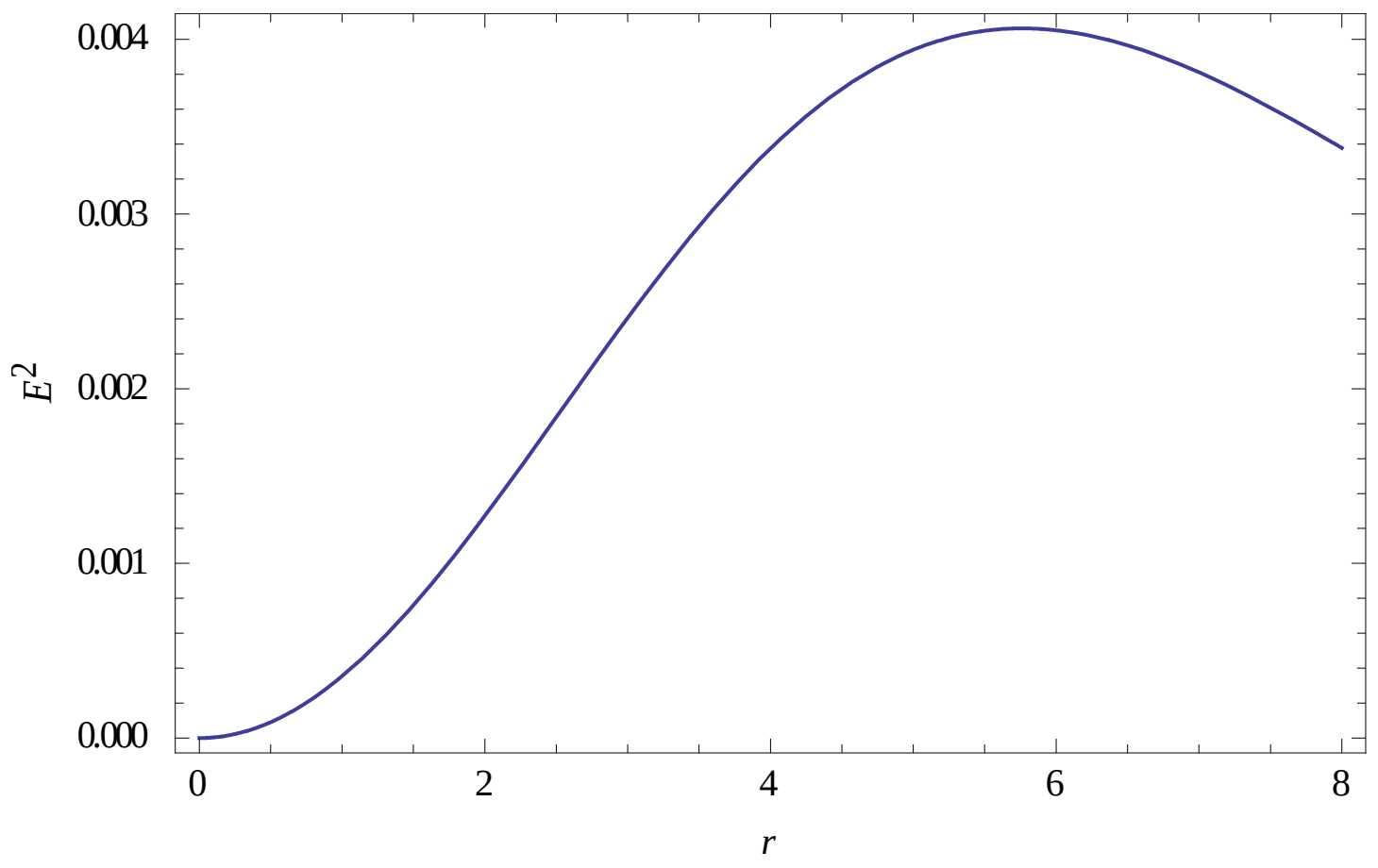

Fig. 5: The variation of the electric field with respect to $r$ by taking $A = 0.0183209$ , $B = 0.0139166673$ and $\frac{M}{R} = 0.299$ .

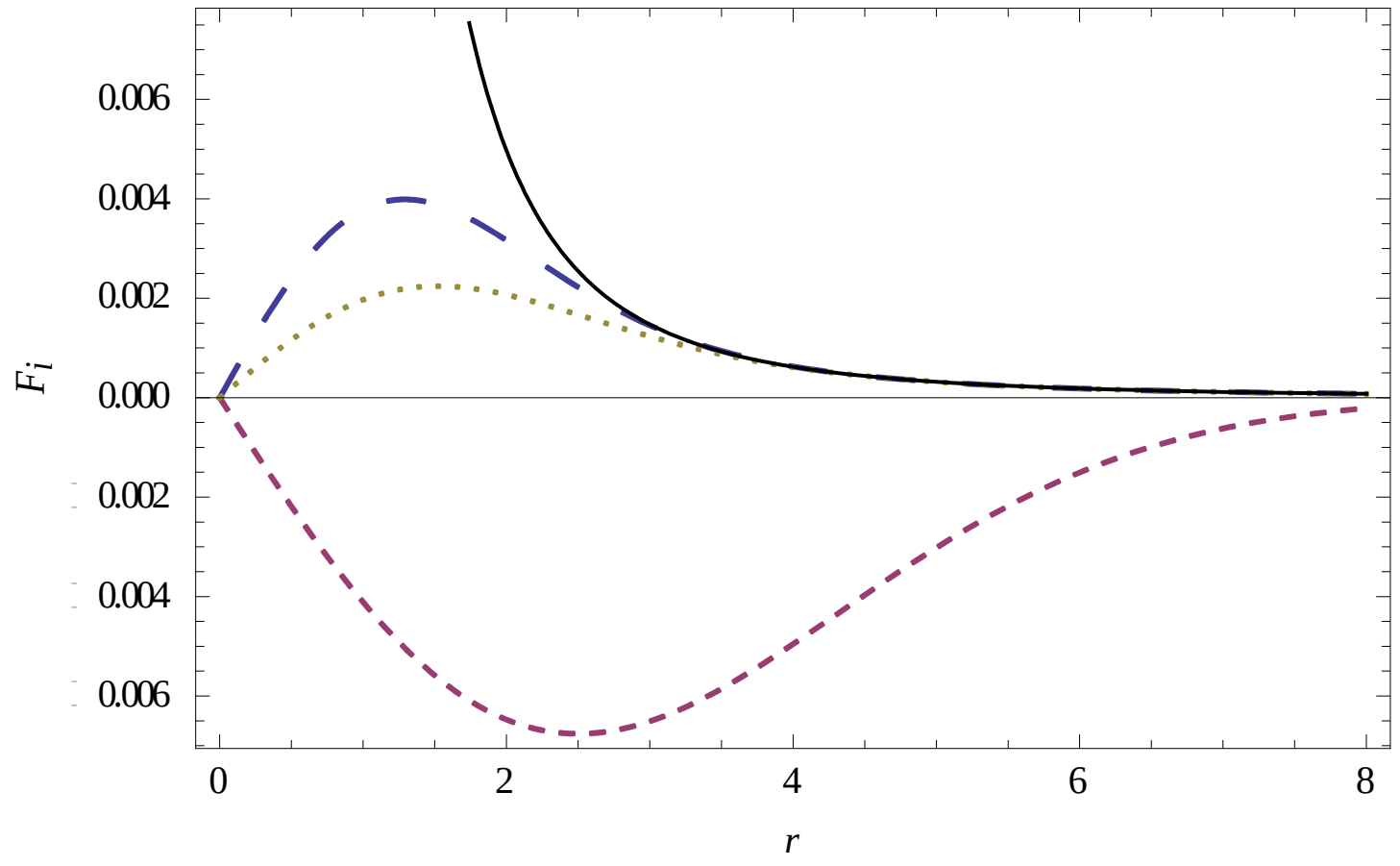

Fig. 6: solid, dashed, dotted and large dashed line represents $F_e$, $F_g$, $F_a$ and $F_h$ respectively against $r$ by taking $A = 0.0183209$, $B = 0.0139166673$ and $\frac{M}{R} = 0.299$.

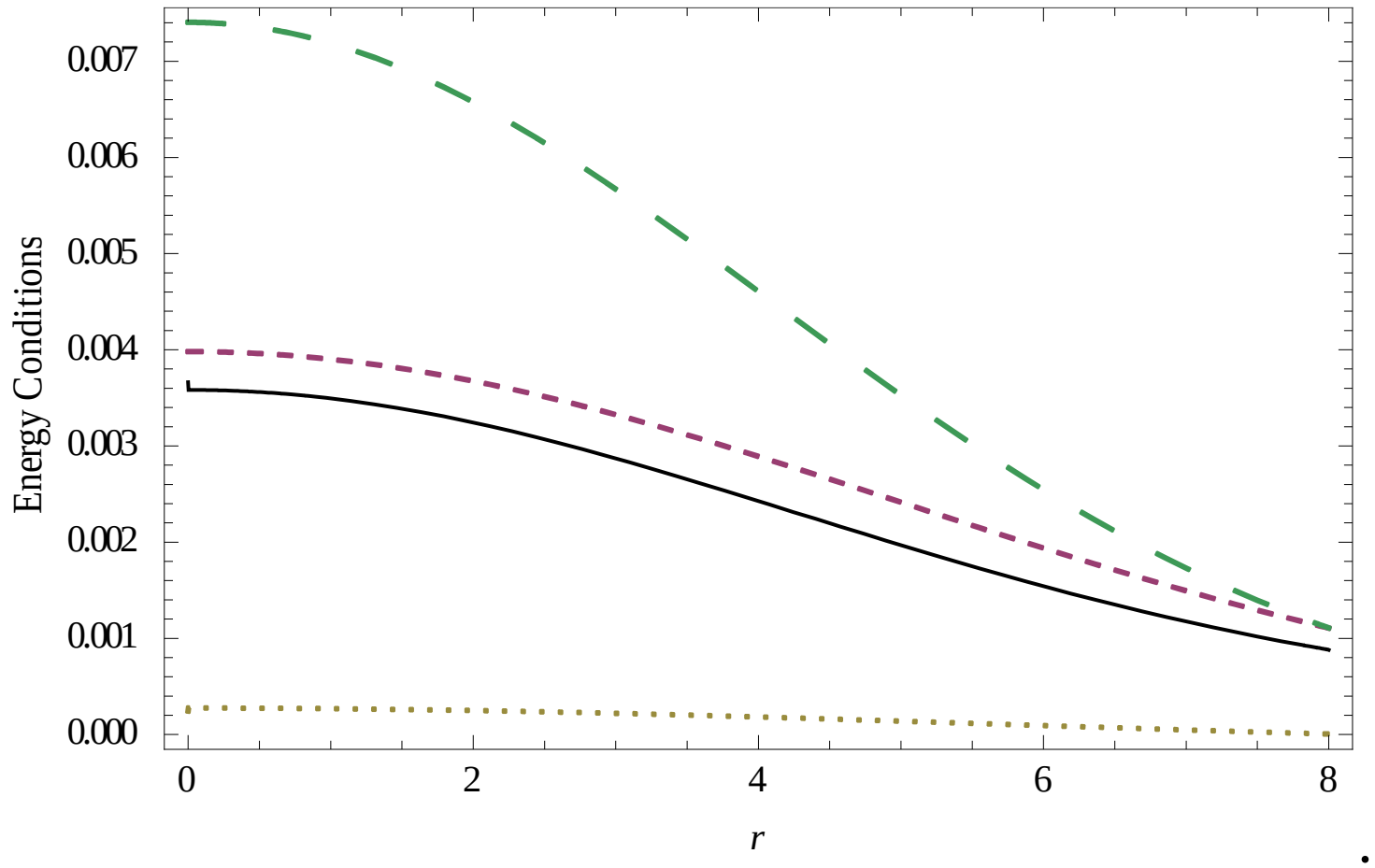

.

Fig. 7: solid, dashed, dotted and large dashed line represents $NEC$, $WEC_r$, $WEC_t$ and $SEC$ respectively against $r$ by taking $A = 0.0183209$ , $B = 0.013916673$ and $\frac{M}{R} = 0.299$ .

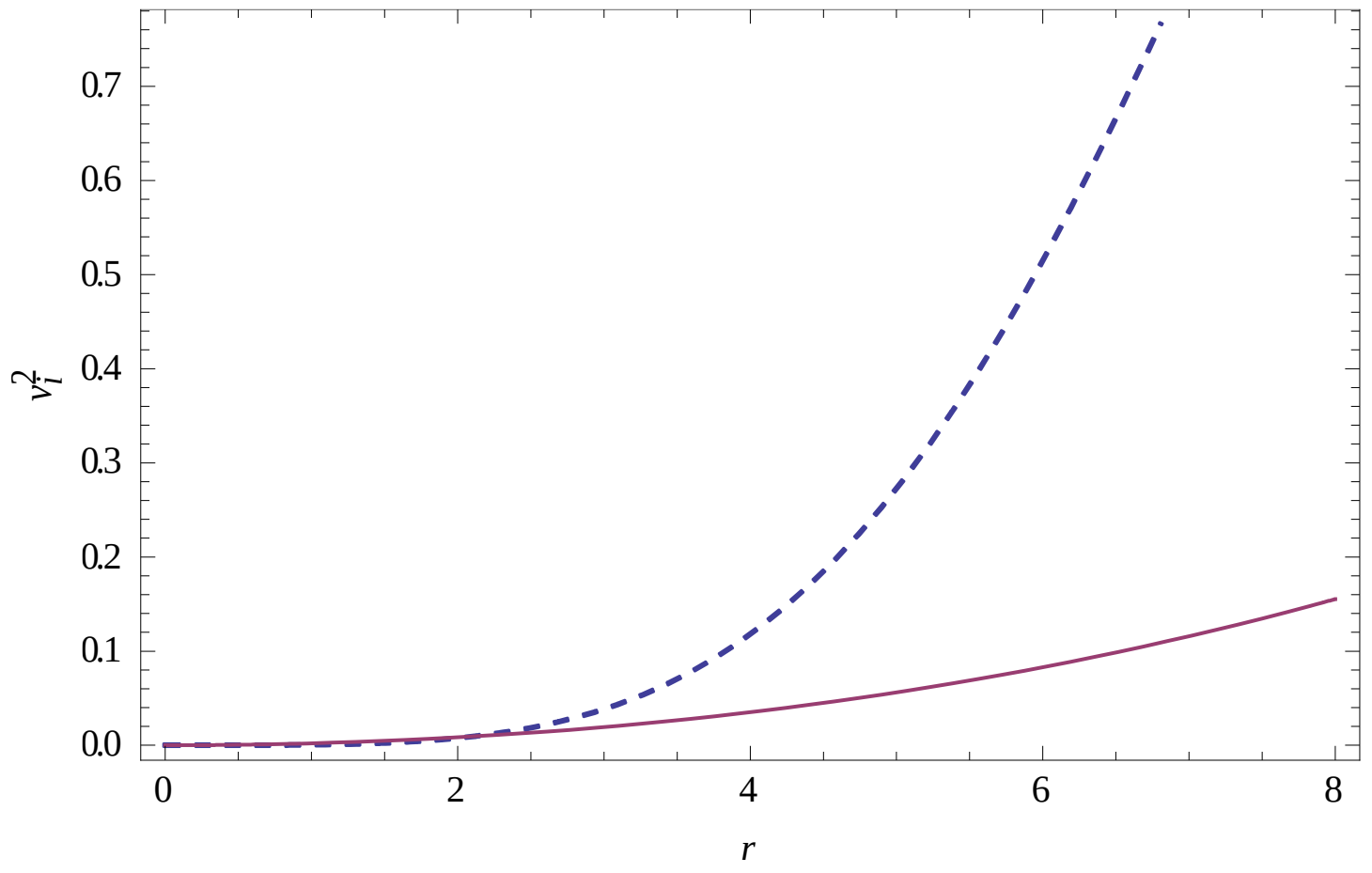

Fig. 8: The dotted and solid line represents radial sound speed $v_{sr}^2$ and tangential sound speed $v_{st}^2$ respectively against $r$ by taking $A = 0.0183209$ , $B = 0.013916673$ and $\frac{M}{R} = 0.299$ .

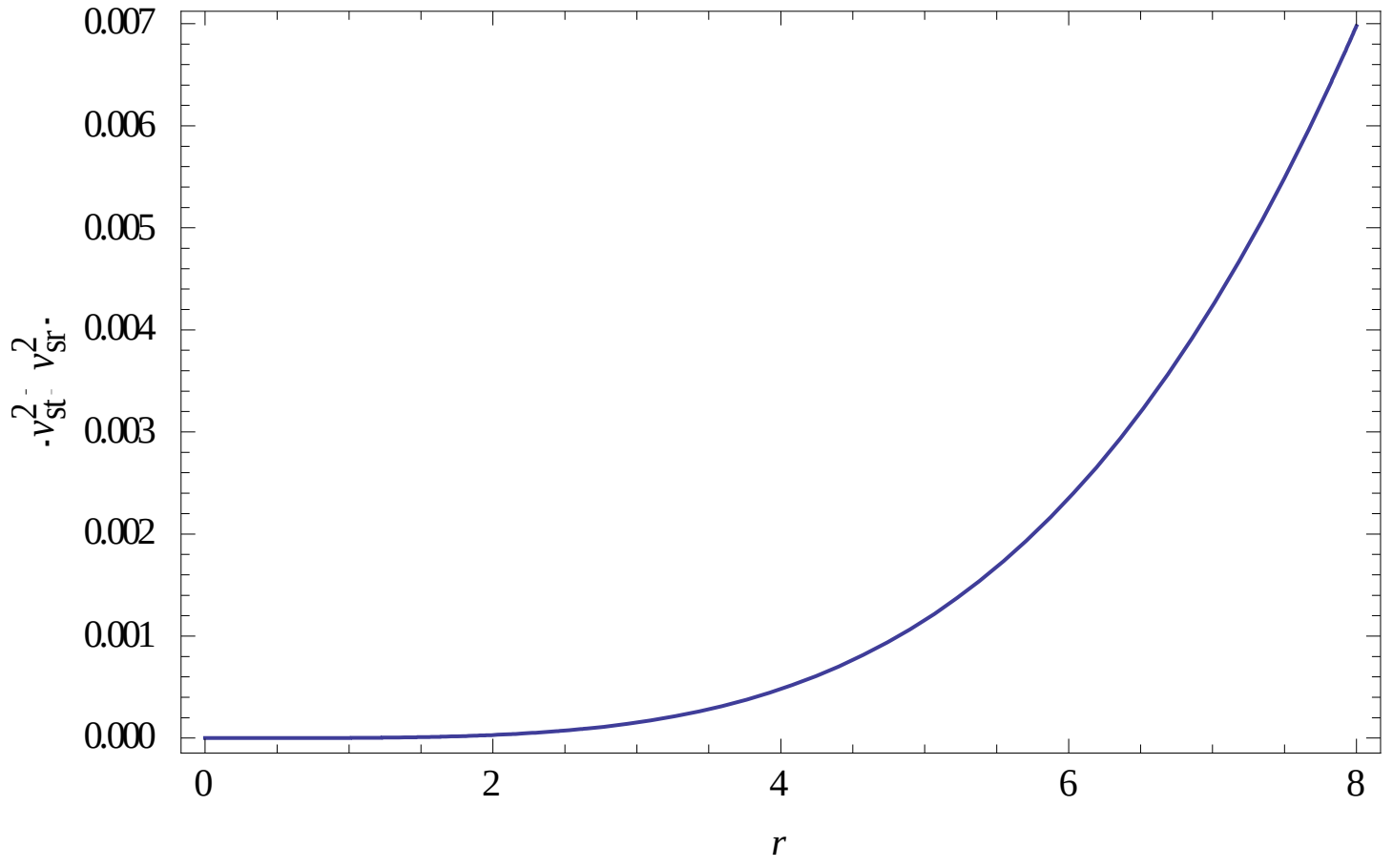

Fig. 9: The variation of $\left| v_{st}^2 - v_{sr}^2 \right|$ is shown against $r$ by taking $A = 0.0183209$ , $B = 0.013916673$ and $\frac{M}{R} = 0.299$ .

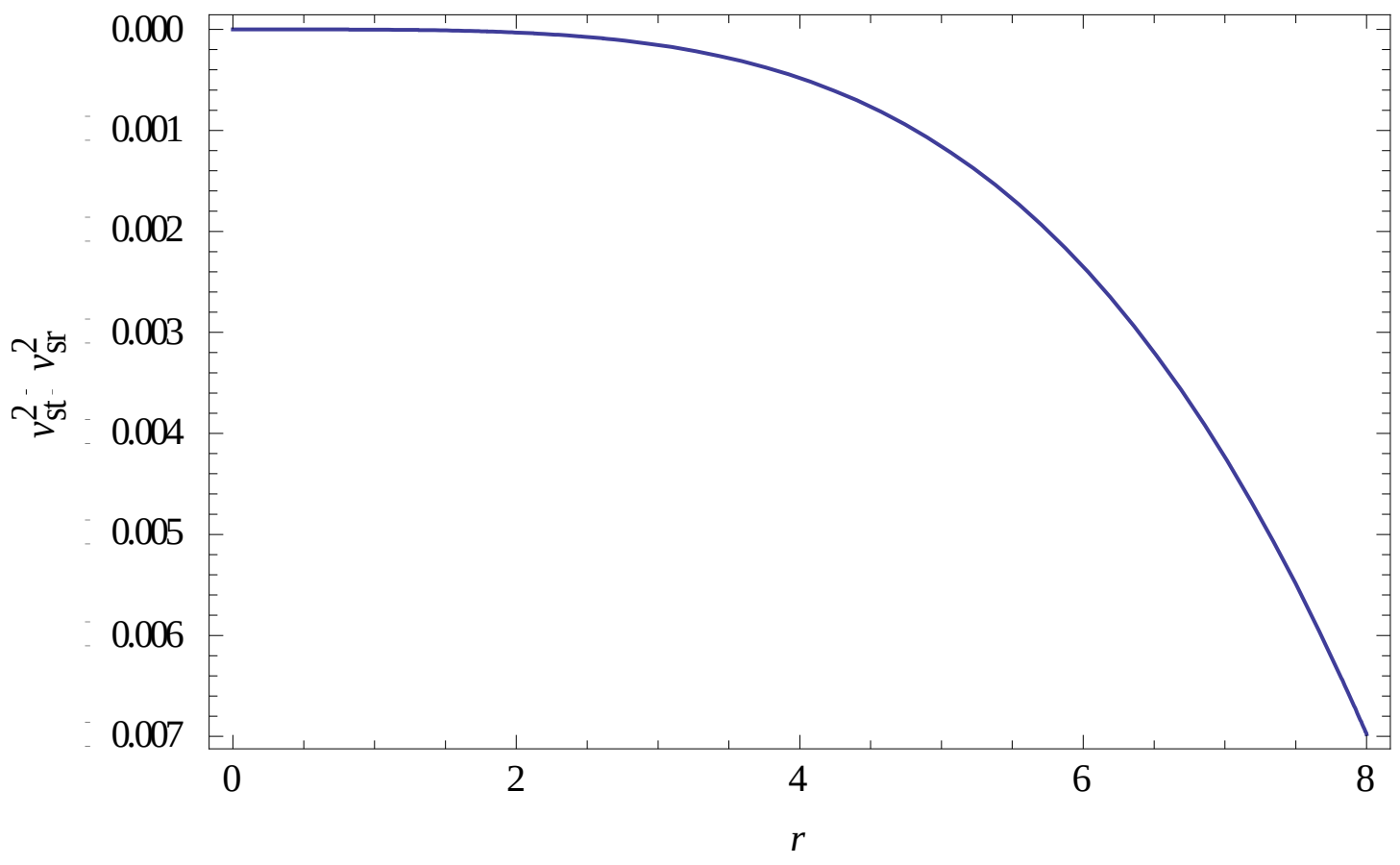

Fig. 10: The variation of $v_{st}^2 - v_{sr}^2$ is shown against $r$ by taking $A = 0.0183209$ , $B = 0.013916673$ and $\frac{M}{R} = 0.299$ .

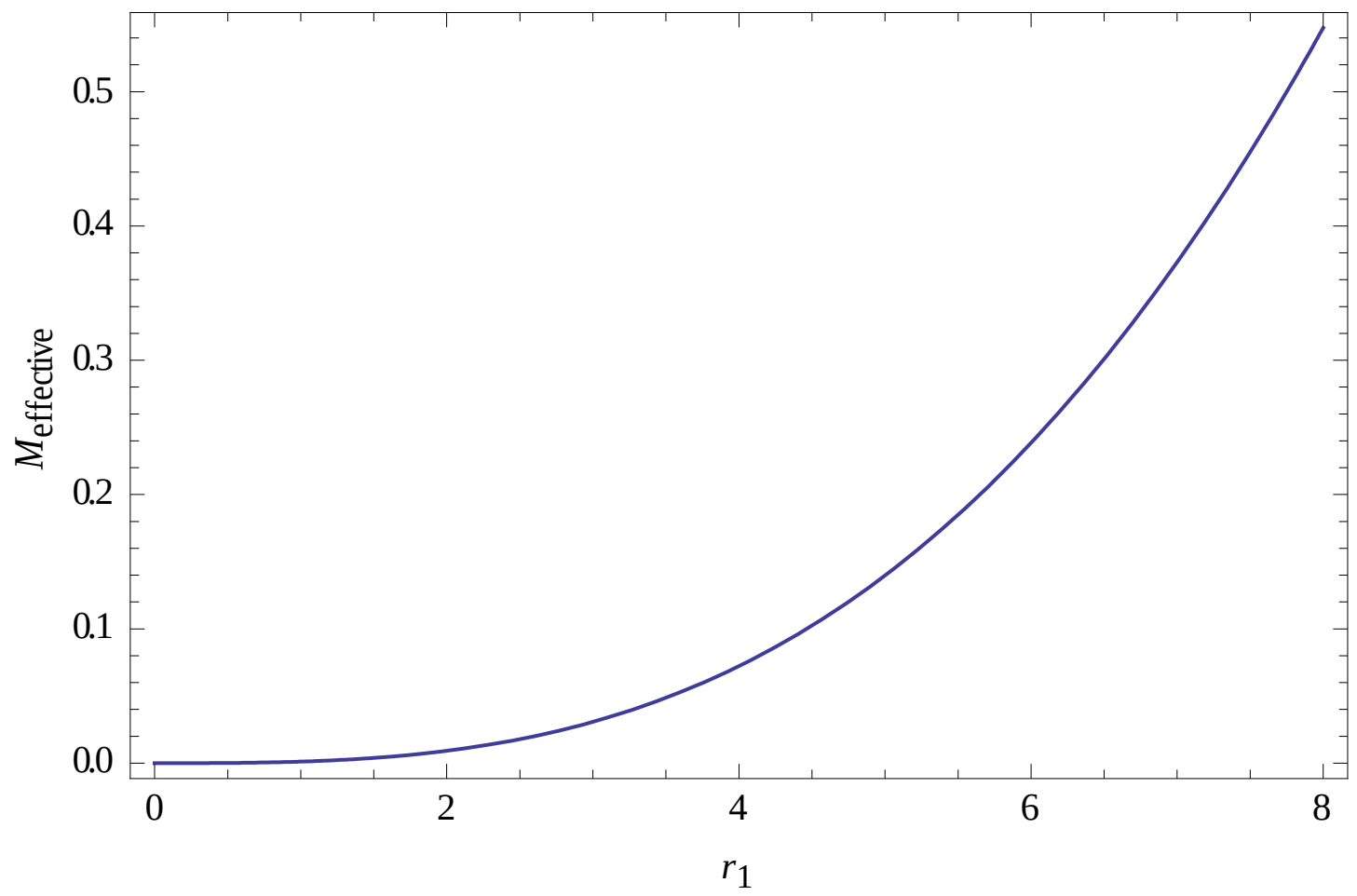

Fig. 11: The variation of $M_{eff}$ is shown against $r_1$ by taking $A = 0.0183209$ , $B = 0.013916673$ and $\frac{M}{R} = 0.299$ .

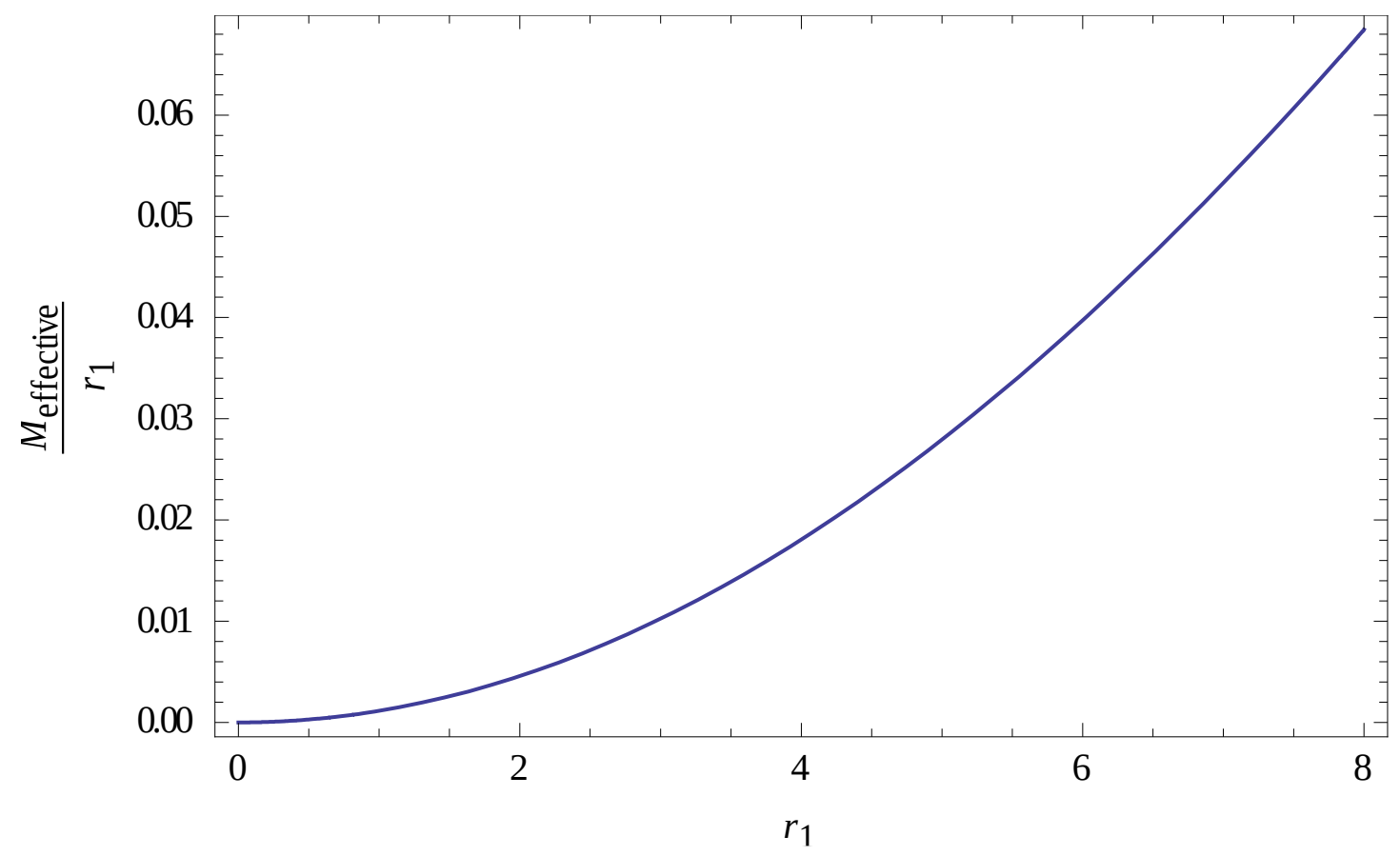

Fig.12: The variation of $\frac{M_{eff}}{r_1}$ is shown against $r_1$ by taking $A = 0.0183209$ , $B = 0.0139166673$ and $\frac{M}{R} = 0.299$ .

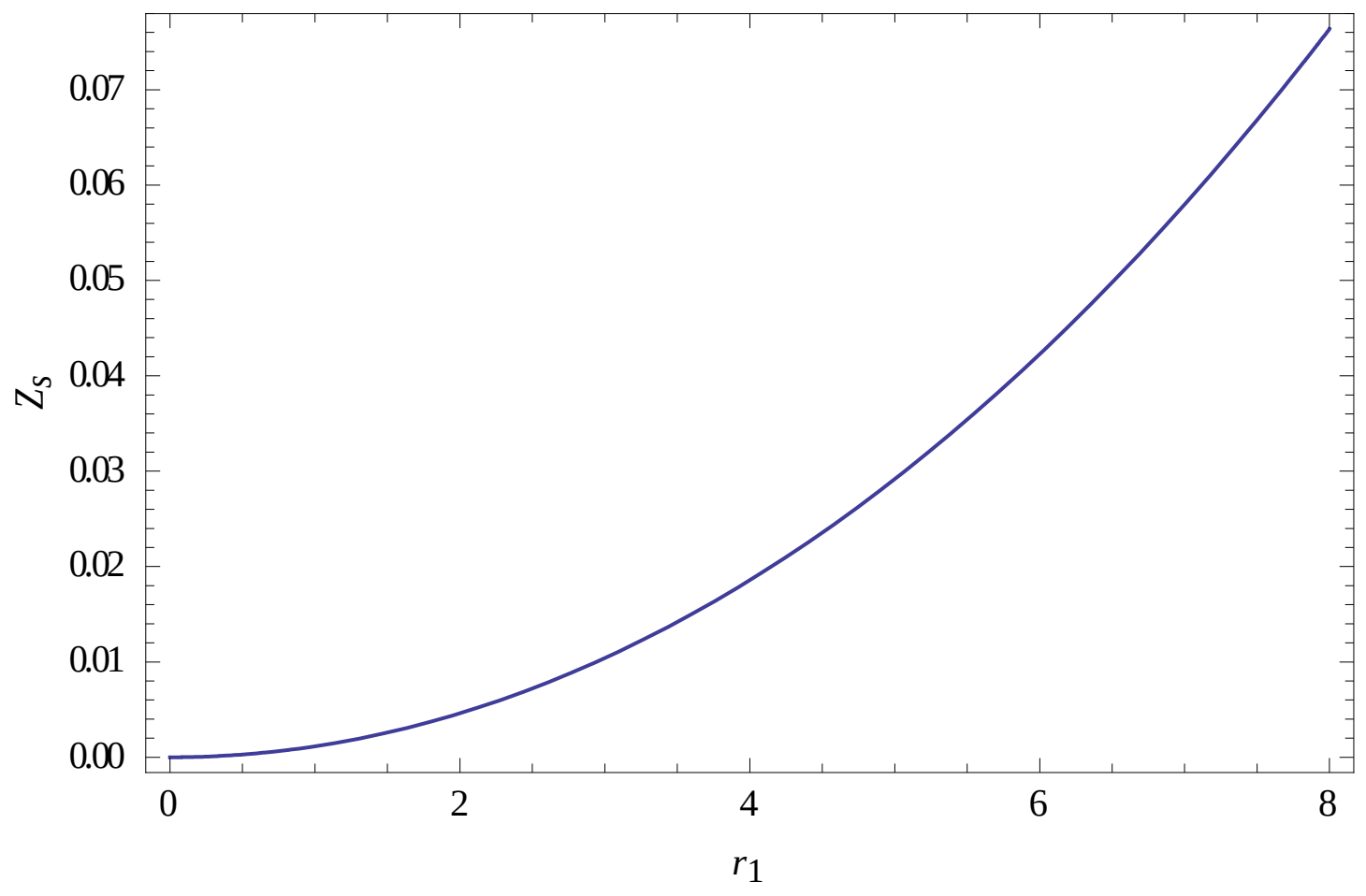

Fig. 13: The variation of redshift function $Z_s$ is shown against $r_1$ by taking $A = 0.0183209$ and $\frac{M}{R} = 0.299$.

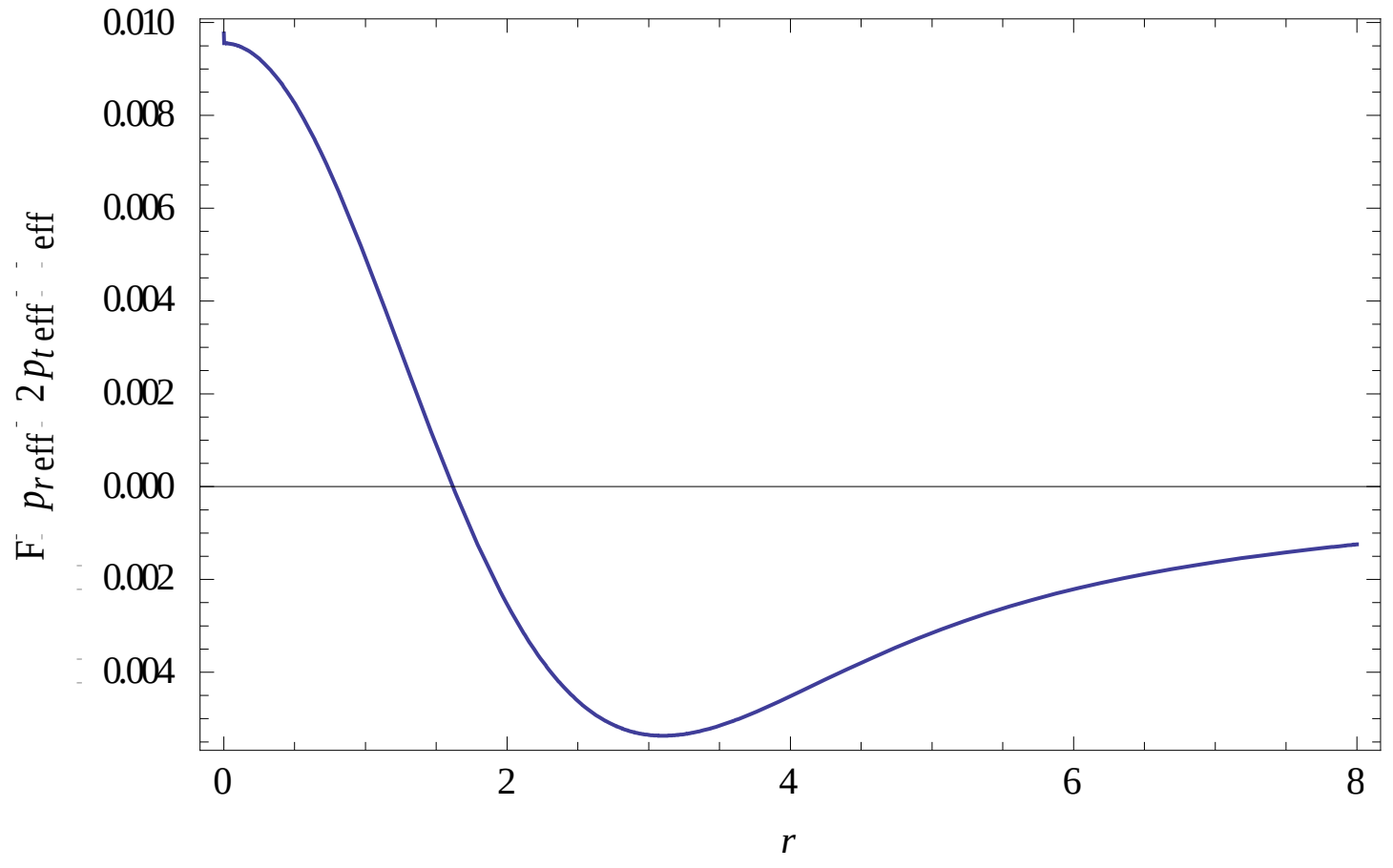

Fig. 14: The variation of $p_{r\,eff} + 2p_{t\,eff} - \rho_{eff}$ is shown against $r$ by taking $A = 0.0183209$, $B = 0.0139166673$ and $\frac{M}{R} = 0.299$.